\documentclass[aps,pra,showpacs,twocolumn,nofootinbib]{revtex4}
\usepackage{amsmath,graphicx,epsfig,amssymb,subfigure,times}
\usepackage[usenames]{color}

\newcommand{\h}[2][ ]{\hat{#2}^{\vphantom{\dag} #1}}
\newcommand{\hd}[2][ ]{\hat{#2}^{\dag #1}}

\begin{document}

\title{Atomic entanglement generation and detection via degenerate four-wave-mixing of a Bose-Einstein condensate in an optical lattice}

\author{Andrew~J.~Ferris}
\author{Murray~K.~Olsen}
\author{Matthew~J.~Davis}
\affiliation{The University of Queensland, School of Mathematics and Physics, ARC Centre of Excellence for Quantum-Atom Optics, Qld 4072, Australia}

\date{\today}
% blah test
\begin{abstract}

The unequivocal detection of entanglement between two distinct matter-wave pulses is a significant challenge that has yet to be experimentally demonstrated.  We describe a realistic scheme to  generate and detect continuous variable entanglement between two atomic matter-wave pulses produced via degenerate four-wave-mixing from an initially trapped Bose-Einstein condensate loaded into a one-dimensional optical lattice.  We perform a comprehensive numerical investigation for fixed condensate parameters to determine the maximum violation of separability and Einstein-Podolsky-Rosen inequalities for field quadrature entanglement, and describe and simulate an experimental scheme for measuring the necessary quadratures.

\end{abstract}


\maketitle

\section{Introduction}

\label{sec_intro}
Recently there has been much interest in the quantum properties of matter waves and the study of quantum atom optics~\cite{Rolston2002,Molmer2003}. The field has grown out of atom optics and encompasses concepts and ideas from quantum optics, condensed matter theory, atomic and molecular physics and quantum information theory.  Experimentally, ultra-cold atoms provide a clean and controllable environment to investigate a wide range of new and existing models. Notable examples include the observation of the BCS--BEC crossover regime~\cite{Greiner2002} and the Mott insulator--superfluid quantum phase transition~\cite{Bloch2008}.

In addition to answering fundamental questions of science, this developing area of physics has the potential for new and exciting applications. It is predicted that quantum entanglement will enable a novel set of technologies based on fundamental quantum principles, such as precision measurement devices~\cite{Giovannetti2004} and quantum computers~\cite{Ekert1996}. The precise coherent quantum manipulation of ultra-cold atomic systems has been demonstrated in many experiments and holds promise for these systems to be candidates for quantum information applications in the future.

There have been a number of demonstrations of quantum atom optical phenomena in recent years. These include the observation of non-classical effects in atomic fields such as the Hanbury Brown-Twiss effect for bosons~\cite{Schellekens2005a,Oettl2005}, anti-bunching for fermions~\cite{Vassen2007a}, reduced local pair correlations~\cite{Tolra2004,Kinoshita2005}, sub-Poissonian number fluctuations~\cite{Chuu2005a,Esteve2008}, and density correlations from molecular disassociation~\cite{Greiner2005a}, atomic collisions~\cite{Perrin2007}, and in the Mott-insulator regime in an optical lattice~\cite{Folling2005a}. The majority of these investigations have concentrated on correlations in the atom number or density. Future applications of quantum atom optics utilizing entanglement and squeezing will require manipulation and detection of the phase of the quantum state of matter-waves. This presents a challenge as creating stable phase references and performing mode-matched intereference is likely to be difficult for ultracold atoms.

Entanglement can be generated and utilized in a variety of forms. In this paper we address the generation and detection of continuous-variable entanglement between orthogonal spatial modes of an atomic field in a second-quantized formalism. This is distinct from entanglement between single particles, as seen in a first-quantized picture. Such continuous-variable entanglement has been extensively used in the field of quantum optics~\cite{Braunstein2005}.

Entangled and squeezed states of the electromagnetic field have been generated  in quantum optics experiments, and proof-of-principle demonstrations for potential applications such as quantum cryptography have been successful~\cite{Braunstein2005}. Photons may not be ideal for all quantum operations; light travels fast but is difficult to contain, and the lack of mass makes it relatively insensitive to rotation and acceleration in interferometric sensors~\cite{Clauser1988}. In principle, the achievements of quantum optics can be replicated with bosonic atoms. Indeed, there have been impressive achievements in entangling the spins of two distinct atomic clouds in the continuous-variable limit~\cite{Julsgaard2001}. Recently, entanglement in the reduced two-body density matrix was demonstrated by number and phase correlations in double- and few-well systems~\cite{Esteve2008}. However, no experiment has yet demonstrated entanglement between the spatial modes of an atomic field.

Several suggestions have emerged for creating entanglement in a Bose-Einstein condensate (BEC). One possibility is to transfer entanglement from an pre-existing entangled source to the atoms. For instance, the state of entangled light can be mapped onto atomic modes by using lasers to induce a Raman transition~\cite{Haine2005a,Olsen2008}. A related suggestion for entanglement detection involves mapping the atom statistics onto photons in the reverse process, in order to access well-developed single photon and quadrature measurement techniques~\cite{Bradley2007a,Olsen2008}.

A second approach is to generate the entanglement directly in the atomic system. Atomic systems exhibit three- and four-wave mixing processes that can generate entanglement in light. The dissociation of a diatomic molecular Bose-Einstein condensates into entangled pairs is analogous to the process of three-wave mixing in an optical parametric amplifier~\cite{Kheruntsyan2005,Kheruntsyan2005b}. In the second-quantized picture the spatial modes of the products become entangled. Unfortunately, the  entanglement that results is difficult to use or detect because of the necessity to use a phase reference that is correlated to the phase of both the atomic and molecular modes.  Atom-light scattering is an example of a four-wave mixing process which can entangle atomic and photonic modes~\cite{Gasenzer2002}. Superradiant Rayleigh scattering occurs in an elongated condensate, where the scattering process is self-stimulating, generating only one or a few macroscopically occupied, entangled modes~\cite{Inouye1999,Schneble2004,vanderStam2007}.

The entanglement we consider in this paper occurs between modes of the atomic field. The inherent s-wave scattering between ultra-cold atoms can result in the four-wave mixing of matter waves. Stimulated four-wave mixing between colliding condensates was first reported in 1999~\cite{Deng1999}, followed by observations of spontaneous four-wave mixing and the demonstration of the potential reversibility of the process~\cite{Vogels2002}. In principle, this four-wave mixing scheme produces many pairs of spatially entangled modes~\cite{Perrin2008,Ogren2008}, in analogy to a recent optical experiment~\cite{Boyer2008}. However, the sheer number of resonant modes limits the gain and entanglement generated between any given pair of modes, as the condensate is rapidly depleted. In this work we extend a suggestion to use \emph{degenerate} four-wave mixing (where the two input waves are the same mode) of a quasi-1D BEC in an optical lattice~\cite{Hilligsoe2005a,Campbell2006a} to create and detect continuous variable entanglement.  The quasi-1D geometry  limits the number of modes for resonant collisions, and leads to the generation two separate yet entangled matter waves.

Hilligs{\o}e and M{\o}lmer~\cite{Hilligsoe2005a} first suggested loading a stationary Bose-Einstein condensate into a moving optical lattice in order to make a degenerate four-wave-mixing process resonant. There is a large body of work on the inherent dynamical instabilities in moving optical lattices which cause heating of BECs (see~\cite{WuNuBattle,Konotop2002,Ferris2008} and references therein). Interestingly, the same dynamical instability can be interpreted as a collisional process that generates entangled modes of specific momenta. An experiment was subsequently performed by Campbell \emph{et al.}~\cite{Campbell2006a} that shows pairs of modes are populated with the momenta predicted by mean-field theory; however there has been no experimental proof of entanglement in this system.  The goal of this work is to propose a method to demonstrate entanglement between modes populated by degenerate four-wave-mixing.

To utilize or detect the presence of entanglement requires a suitable phase reference~\cite{Braunstein2005,Bartlett2005,Bartlett2006} which is provided by the near classical coherent output of a laser in optics.  In atom optics the equivalent of laser light is a Bose-Einstein condensate.  However, in experiments the size of BECs is typically limited to be between $10^3$ and $10^8$ atoms, and interactions between atoms result in atom losses, number dependent phase evolution, and phase diffusion meaning that they are less than ideal as a phase reference.  Also, setting aside a separate condensate of atoms to use as a phase reference may not be practical experimentally.

In Ref.~\cite{Ferris2008b} we extended the separability criterion of Duan \emph{et al.}~\cite{Duan2000a} and Simon~\cite{Simon2000a}, and the criterion for demonstration of the Einstein-Podolsky-Rosen (EPR) paradox of Reid~\cite{Reid1989a} to situations where no assumptions could be made about the quantum state of the available phase reference.  This allows the possibility of using a non-classical local oscillator in balanced homodyne measurement schemes for quadrature measurements that are common in quantum optics.  We then applied these criteria to a toy four mode-model of degenerate four-wave mixing in a Bose-Einstein condensate with up to 1000 atoms, and showed that appropriate beam-splitting operations combined with atom counting could be used to demonstrate both inseparability and the EPR paradox.

In this paper we extend this approach to perform one-dimensional simulations of a more realistic experimental system.  We simulate the adiabatic loading of a trapped Bose-Einstein condensate into a moving optical lattice to initiate degenerate four-wave mixing, before turning off the lattice potential and performing beam-splitting operations with appropriate Bragg pulses to generate four distinct atomic clouds.  Performing number difference measurements between these clouds provides access to the atomic quadratures and the subsequent violation of the entanglement inequalities.  We determine how the measure of entanglement depends on the time held in the optical lattice, the relative phase of the Bragg pulses, and the number of seed atoms in the outgoing modes.  We conclude that these experimental complications will not necessarily prevent the demonstration of entanglement of matter-waves in such a system.

The paper is organized as follows. In Sec.~\ref{sec_d4wm} we review the degenerate four-wave mixing of a BEC in an optical lattice that was the subject of previous theoretical and experimental investigations~\cite{Hilligsoe2005a,Campbell2006a,Olsen2006a}. Section~\ref{sec_scheme} details the proposed measurement scheme, including the basics of Bragg pulses, homodyne measurements and the entanglement criteria we employ. The results of numerically implementing our scheme are presented in Sec.~\ref{sec_results} before we conclude in Sec.~\ref{sec_conclusion}.

\section{Degenerate four-wave mixing of matter-waves}

\label{sec_d4wm}

The dominant interaction between ultra-cold atoms in a Bose-Einstein condensate is s-wave scattering. In free space the conservation of energy and momentum in the collision of two particles requires that the outgoing particles have momenta falling on two opposing points of a sphere in $k$-space in the centre of mass frame.  Such s-wave scattering spheres have been observed by colliding two BECs~\cite{Vogels2002,Perrin2007}. Alternatively, this can be viewed as a four-wave mixing process which is resonant and phase-matched on the s-wave scattering sphere.

The presence of a periodic potential alters the dispersion relation for the particles and thus the resonance conditions for collisions. Hilligs{\o}e and M{\o}lmer~\cite{Hilligsoe2005a} showed that adiabatically loading a BEC into a moving optical lattice can allow degenerate four-wave mixing, where the two incident particles are in the same initial state. The optical lattice is created by shining two lasers with wave-vectors $\mathbf{k}_1$ and $\mathbf{k}_2$ of similar frequencies $\omega_1$ and $\omega_2$ onto the atomic cloud. The light creates an effective potential for the atoms
\begin{equation}
  V(\mathbf{r}) = \frac{V_L}{2} \sin(2\mathbf{k}_L \cdot \mathbf{r} - 
  \delta t + \theta), \label{lattice} % Check same as code
\end{equation}
where $2\mathbf{k}_L = \mathbf{k}_1 - \mathbf{k}_2$, $\delta = \omega_1 - \omega_2$, $\theta$ is the relative phase between the beams, and $V_L$ is  proportional to the intensity of the lasers and depends on the details of the atom-light interaction. It is convenient to write $V_L = sE_R$ where $s$ characterizes the strength of the optical lattice, and $E_R = \hbar^2 k_L^2/2m$ is the recoil energy.

\begin{figure}[t]
\begin{centering}
\includegraphics[width=7.5cm]{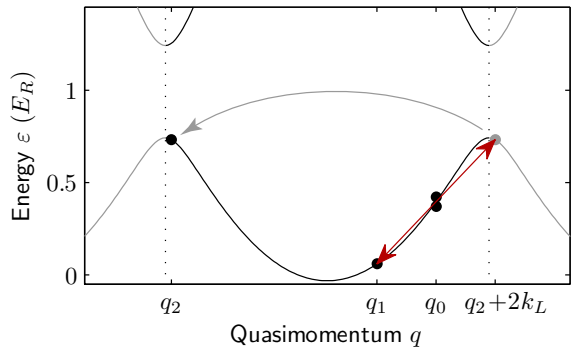}
\caption{(Color online) Band structure of an optical lattice with $s = 1$. Two atoms in the (moving) condensate mode 0 can collide into modes 1 and 2, conserving energy and quasi-momentum.}
\label{fig_bands}
\end{centering}
\end{figure}

The degenerate four-wave-mixing process is illustrated in Fig.~\ref{fig_bands} where two particles in mode 0 with quasimomentum $q_0$ in the lowest Bloch band can collide, transferring an atom to each of modes 1 and 2 with quasimomenta $q_1$ and $q_2$. During this process they conserve their total energy $2\varepsilon_0 = \varepsilon_1 + \varepsilon_2$ and quasimomentum $2q_0 = q_1 + q_2$~modulo~$2k_L$.  By subsequently adiabatically turning off the optical lattice the generated quasi-momentum states are converted to momentum states in free space between $\pm k_L$. If we write resulting momentum of mode 1 as $k_1 = k_0 - \Delta k$, then the momentum of mode 2 is $k_2 = k_0 + \Delta k - 2k_L$.

Defining the annihilation operator of mode $j$ to be $\h{a}_j$ the model Hamiltonian describing this process can be written as
\begin{equation}
 \h{H} = i\hbar\chi\left( \h[2]{a}_0\hd{a}_1\hd{a}_2 - \hd[2]{a}_0\h{a}_1\h{a}_2 \right). \label{simple_hamiltonian}
\end{equation}
This pairwise scattering process is predicted to produce sub-Poissonian number correlations and quadrature entanglement between the atoms with quasimomenta $q_1$ and $q_2$~\cite{Hilligsoe2005a,Olsen2006a}. However, the measurement of number correlations or demonstration of entanglement was not accomplished in the sole experimental realisation of degenerate four-wave mixing in a BEC to date~\cite{Campbell2006a}.

We can perform a simplified analysis of the Hamiltonian in Eq. (\ref{simple_hamiltonian}) using the undepleted pump approximation where we assume that mode 0 begins in a large coherent state $|\alpha\rangle$, where $\alpha$ is real and the number of particles $N_0 = \alpha^2$. At small enough times one can assume the population of mode 0 is undepleted and therefore the Hamiltonian can be approximated as
\begin{equation}
 \h{H} \approx i\hbar N_0\chi\left( \hd{a}_1\hd{a}_2 - \h{a}_1\h{a}_2 \right). \label{upa}
\end{equation}
In the Heisenberg picture the solution for this system is~\cite{Braunstein2005,Olsen2006a}
\begin{gather}
  \h{a}_1(t) = \cosh(N_0\chi t)\h{a}_1(0) + \sinh(N_0\chi t)\hd{a}_2(0), \nonumber \\
  \h{a}_2(t) = \cosh(N_0\chi t)\h{a}_2(0) + \sinh(N_0\chi t)\hd{a}_1(0). \label{upa_solution}
\end{gather}
If modes 1 and 2 are initially vacuum then they make up a two-mode squeezed state that is well-known from quantum optics~\cite{WallsMilburn}. The two modes are exactly correlated in number, and anti-correlated in phase. Measuring the correlation between the number of atoms in modes 1 and 2 is, however, insufficient to prove entanglement, as it is possible to construct a separable density matrix that is consistent with any set of local number measurement outcomes~\cite{Bartlett2006}.

To unequivocally demonstrate that entanglement exists between the two modes it is necessary to perform phase sensitive measurements. The phase difference between the modes is insensitive to anti-correlations, and thus measurement of the \emph{relative} phase will not help demonstrate entanglement between these states. We therefore need to measure the phase quadratures of both modes.

Quadrature measurements have been used to prove entanglement between photonic modes in quantum optics experiments~\cite{Braunstein2005}. A third phase reference, or local oscillator, is interfered with the entangled modes on a beam splitter to reveal the quantum correlations. A large and coherent local oscillator (produced by a laser) allows one to make accurate measurements of the field quadratures, defined as
\begin{equation}
  \h{X}_j = \h{a}_j + \hd{a}_j, \quad \quad \h{Y}_j = i\bigl(\h{a}_j - \hd{a}_j\bigr). \label{simple_quads}
\end{equation}
Duan \emph{et al.}~\cite{Duan2000a} and Simon~\cite{Simon2000a} have derived a simple criteria for the separability of the two modes. All separable states obey
\begin{equation}
  \mathrm{Var}\bigl[\h{X}_1 - \h{X}_2\bigr] + \mathrm{Var}\bigl[\h{Y}_1 + \h{Y}_2\bigr] \ge 4, \label{Duan}
\end{equation}
where we write the variance $\mathrm{Var}\bigl[\h{A}\bigr] = \langle\h[2]{A}\rangle - \langle\h{A}\rangle^2$. Violation of the above inequality indicates that the system must be entangled. For all times $t>0$ the solution Eq.~(\ref{upa_solution}) of the Hamiltonian~(\ref{upa}) in the undepleted pump approximation violates this bound
\begin{equation}
 \mathrm{Var}\bigl[\h{X}_1 - \h{X}_2\bigr] + \mathrm{Var}\bigl[\h{Y}_1 + \h{Y}_2\bigr] = 4e^{-2N_0\chi t}.
\end{equation}
This inequality is useful for detecting the entanglement generated by the four-wave mixing process, and has been successfully employed in optical experiments~\cite{Braunstein2005}. The Einstein-Podolsky-Rosen paradox has been demonstrameted in optical fields using similar inequalities~\cite{Reid1989a,Braunstein2005}.

\section{Entanglement detection}

\label{sec_scheme}

In optics the quadrature measurements required to demonstrate entanglement can be made by balanced homodyning that mixes the signal beams with a phase reference beam on a 50-50 beam splitter. We employ balanced homodyning for atoms as this method typically has a superior signal-to-noise ratio compared to unbalanced homodyne schemes. In the Heisenberg picture we define $\h{a} \equiv \h{a}_{\mathrm{in}}$ and $\h{b} \equiv \h{b}_{\mathrm{in}}$ as the annihilation operators for the signal and local oscillator before the beam splitting. The output modes $\h{a}_{\mathrm{out}}$ and $\h{b}_{\mathrm{out}}$ are given by
\begin{equation} \label{beamsplit}
 \h{a}_{\mathrm{out}} = (\h{a}_{\mathrm{in}} + \h{b}_{\mathrm{in}})/\sqrt{2}, \hspace{0.25cm}\mathrm{and} \hspace{0.25cm} \h{b}_{\mathrm{out}} = (\h{a}_{\mathrm{in}} - \h{b}_{\mathrm{in}})/\sqrt{2}.
\end{equation}
The final step is to measure the difference in the number of particles exiting the beam splitter ports $\h{a}_{\mathrm{out}}$ and $\h{b}_{\mathrm{out}}$. The measured quadrature is rescaled by the size of the local oscillator, according to
\begin{equation} \label{quad}
  \h{X} = \frac{\hd{a}_{\mathrm{out}} \h{a}_{\mathrm{out}} -
\hd{b}_{\mathrm{out}}\h{b}_{\mathrm{out}}}{\langle \hd{b}_{\mathrm{in}} \h{b}_{\mathrm{in}}\rangle^{1/2}} =
\frac{\hd{a}\h{b} + \h{a}\hd{b}}{\langle \hd{b} \h{b}\rangle^{1/2}} \approx \hd{a} + \h{a},
\end{equation}
where the final approximation is good for phase references that are large coherent states.
% Unlike the photonic counterparts, large sources of atoms to use as local oscillators are not available to perform homodyne measurements.

%In previous work~\cite{Ferris2008b} we have demonstrated that for degenerate four-wave-mixing described by \ref{simple_hamiltonian}, the bosons remaining  in mode 0 are a suitable phase reference for entanglement demonstration for a range of qua.

Thus three ingredients are required to perform homodyne measurements with Bose-Einstein condensates: a suitable phase reference, the atomic equivalent of a beam splitter, and detectors able to accurately count the number or density of atoms in the various modes. Bragg scattering can be used to interfere (beam split) atoms of different momenta, and reasonably efficient atom detection has been demonstrated with multi-channel plate detectors for metastable Helium~\cite{Schellekens2005a,Vassen2007a,Perrin2007}, ionized $^{87}$Rb detection~\cite{Stibor2007}, high numerical aperture optical techniques~\cite{Wilzbach2008}, and utilizing optical cavities~\cite{Oettl2005}. Combined with time-of-flight expansion, the population distribution in momentum space can be directly measured.

\subsection{Bragg pulses}

Bragg scattering is  the process of applying a moving optical lattice to coherently transfer atomic populations from one momentum state to another. The potential in Eq.~(\ref{lattice}) will resonantly transfer an atomic population with momentum $\mathbf{k}_1$ to $\mathbf{k}_2$ (and vice-versa) if and only if $\mathbf{k}_1 - \mathbf{k}_2 = \pm 2\mathbf{k}_L$ and $\delta = \hbar \mathbf{k}_L\cdot(\mathbf{k}_1+\mathbf{k}_2)/m$; this is equivalent to assuming the atom scatters a photon from one beam to the other, conserving both energy and momentum. Higher order scattering is possible, but will not occur if the lattice is relatively weak ($s\lesssim 1$).

A weak lattice held for a duration $\tau = \pi\hbar/sE_R$ provides a $\pi/2$ Bragg pulse, and is equivalent to a 50-50 beam splitter operation. This can be most easily seen in the Heisenberg picture
\begin{eqnarray}
  \h{a}_1(t+\tau) &=& \exp(-i\pi k_1^2/k_L^2 s) \frac{\h{a}_1(t) + \h{a}_2(t)e^{i\theta}}{\sqrt{2}},  \\
  \h{a}_2(t+\tau) &=& \exp(-i\pi k_2^2/k_L^2 s) \frac{\h{a}_2(t) - \h{a}_1(t)e^{-i\theta}}{\sqrt{2}},
\end{eqnarray}
where the pulse begins at time $t$. Note that the relative phase of the two laser beams $\theta$ contributes to the evolution, and this is important when considering quadrature measurements.

In this paper we will make use of mode 0 as a local oscillator in order to detect entanglement between modes 1 and 2.  As simultaneous quadrature measurements of modes 1 and 2 are necessary, we require two local oscillators for balanced homodyning.  This is achieved by first using a $\pi/2$ Bragg pulse to transfer half of the atoms from mode 0 to mode 3 which is initially unpopulated. A weak Bragg pulse will not affect modes 1 or 2 due to the Doppler shift of light.

\subsection{Homodyne measurements}

The final step of the experiment is make a homodyne measurement on mode 1 using mode 0 as the local oscillator, and similarly on mode 2 using mode 3 as the local oscillator. One can simultaneously apply two sets of Bragg pulses tuned to mix momenta $k_1$ with $k_0$, and $k_2$ with $k_3$.

% In Fig.~\ref{fig_pulses} and throughout this paper, we describe the duration of the two pulses as $\tau_H$, ending at time $t_3$.

 To complete the homodyne measurements the population in each of the modes must be measured.
 %At this point the cloud can be allowed to expand in order to produce a time of flight image of the original momentum components.
 The quadrature for mode 1 just before the Bragg pulse is proportional to the \emph{difference} in population found in modes 1 and 0 after the pulse.
\begin{eqnarray}
  \h[\theta_1]{X}_1  & \equiv & \frac{\h{a}_1(t_2)\hd{a}_0(t_2) e^{-i\theta_1} + \hd{a}_1(t_2)\h{a}_0(t_2) e^{i\theta_1}}{\langle\hd{a}_0(t_2)\h{a}_0(t_2)\rangle^{1/2}} \nonumber \\
  & = & \frac{\hd{a}_1(t_3)\h{a}_1(t_3) - \hd{a}_0(t_3)\h{a}_0(t_3)}{\langle\hd{a}_0(t_2)\h{a}_0(t_2)\rangle^{1/2}}, \label{quad1}
\end{eqnarray}
where we have rescaled by the size of the local oscillator. Similarly for modes 2 and 3,
\begin{eqnarray}
  \h[-\theta_2]{X}_2 & \equiv & \frac{\h{a}_2(t_2)\hd{a}_3(t_2) e^{i\theta_2} + \hd{a}_2(t_2)\h{a}_3(t_2) e^{-i\theta_2}}{\langle\hd{a}_3(t_2)\h{a}_3(t_2)\rangle^{1/2}} \nonumber \\
  & = & \frac{\hd{a}_3(t_3)\h{a}_3(t_3) - \hd{a}_2(t_3)\h{a}_2(t_3)}{\langle\hd{a}_3(t_2)\h{a}_3(t_2)\rangle^{1/2}}, \label{quad2}
\end{eqnarray}
As can be seen from the above equation, the $i$th quadrature angle depends on $\theta_i$, the relative phase between the relevant Bragg lasers. The phase is reversed in Eq.~(\ref{quad2}) because $k_2 - k_3$ is negative. The dependence on the relative phase has two important consequences. Firstly, this parameter must be fixed from shot-to-shot in an experiment in order to measure the correct statistics of $\h[\phi]{X}_i$. Secondly, controlling $\theta_1$ and $\theta_2$ allows one to access quadratures of any angle in order to produce a set of measurements that demonstrate entanglement or the EPR paradox.

\subsection{Entanglement criteria}

\label{sec_criteria}

Measurements of the quadrature statistics can confirm the system is entangled by employing the appropriate separability or EPR criteria. In previous work~\cite{Ferris2008b}, we derived three entanglement criteria that take into account the quantum nature of the local oscillator. Earlier studies~\cite{Duan2000a,Simon2000a,Reid1989a} were based on the simplified quadrature operators in Eq.~(\ref{simple_quads}), which in a quantum optics setting closely correspond to the measured quadratures described by Eqs.~(\ref{quad1}) and (\ref{quad2}). However, for the limited numbers of atoms employed in Bose-Einstein condensate experiments and the non-classical state of the local oscillators generated by our proposed scheme, the difference between the simplified and measured quadrature operators is potentially important. The difference between earlier works and the criteria employed below is a direct consequence of the commutation relations of the \emph{measured} quadrature operators,
\begin{gather}
  \left\langle \left[ \h[\phi]{X}_1, \h[\phi]{Y}_1 \right]\right\rangle = 2i \left(1 - \frac{\langle \hd{a}_1\h{a}_1 \rangle }{ \langle \hd{a}_0 \h{a}_0 \rangle } \right),   \nonumber \\
  \left\langle \left[ \h[\phi]{X}_2, \h[\phi]{Y}_2 \right]\right\rangle = 2i \left(1 - \frac{\langle \hd{a}_2\h{a}_2 \rangle }{ \langle \hd{a}_3 \h{a}_3 \rangle } \right), \label{commutator}
\end{gather}
where $\h[\phi]{Y}_i = \h[\phi+\pi/2]{X}_i$.

From this commutator, it follows that all separable states obey~\cite{Ferris2008b}
\begin{multline} \label{Duan2}
  \mathrm{Var}\bigl[\h[\phi_1]{X}_1 - \h[\phi_2]{X}_2\bigr] + \mathrm{Var}\bigl[\h[\phi_1]{Y}_1 + \h[\phi_2]{Y}_2\bigr]
  \ge \\ \;2\left| 1 - \frac{\langle\hd{a}_1
\h{a}_1\rangle}{\langle\hd{b}_1\h{b}_1\rangle} \right|
  + 2 \left| 1 - \frac{\langle\hd{a}_2
\h{a}_2\rangle}{\langle\hd{b}_2\h{b}_2\rangle} \right|.
\end{multline}
Violating this inequality proves the system is entangled. No state can violate the above inequality for all values of $\phi_1$ and $\phi_2$; generally these are experimentally adjusted to find the minimum value (i.e. maximal violation). We can express the left-hand side of Eq.~(\ref{Duan2}) in the form
\begin{eqnarray}
\mathrm{LHS} &=&  \cos^2(\bar{\phi}) \left(\mathrm{Var}\bigl[\h{X}_1 - \h{X}_2\bigr] + \mathrm{Var}\bigl[\h{Y}_1 + \h{Y}_2\bigr] \right) \nonumber \\
 & +& 4\cos(\bar{\phi})\sin(\bar{\phi}) \left( \mathrm{Var}\bigl[\h{X}_1, \h{Y}_2\bigr] +  \mathrm{Var}\bigl[\h{Y}_1, \h{X}_2\bigr]\right) \nonumber\\
 & + &\sin^2(\bar{\phi}) \left(\mathrm{Var}\bigl[\h{X}_1 + \h{X}_2\bigr] + \mathrm{Var}\bigl[\h{Y}_1 - \h{Y}_2\bigr] \right), \label{sep_angle}
\end{eqnarray}
where the omitted superscript implies a quadrature angle of 0, $\bar{\phi} = (\phi_1 + \phi_2)/2$ and $\mathrm{Var}\bigl[\h{A},\h{B}\bigr] = \frac{1}{2}\langle\h{A}\h{B}\rangle + \frac{1}{2}\langle\h{B}\h{A}\rangle - \langle\h{A}\rangle\langle\h{B}\rangle$ is the covariance of $\h{A}$ and $\h{B}$. Note that only the sum of the quadrature angles enters the expression; the value is independent of the difference $\phi_1 - \phi_2$. This simplifies the minimization problem to just one variable with a simple sinusoidal form.

For brevity, we define the separability parameter $\mathcal{S}$,
\begin{equation}
  \mathcal{S} = 2\frac{\mathrm{Var}\bigl[\h[\phi_1]{X}_1 - \h[\phi_2]{X}_2\bigr] + \mathrm{Var}\bigl[\h[\phi_1]{Y}_1 + \h[\phi_2]{Y}_2\bigr]}{\left| 1 - \langle\hd{a}_1
\h{a}_1\rangle/\langle\hd{b}_1\h{b}_1\rangle \right|
  + \left| 1 - \langle\hd{a}_2
\h{a}_2\rangle/\langle\hd{b}_2\h{b}_2\rangle \right|}. \label{sep_param}
\end{equation}
The separability criteria is then simply $\mathcal{S} \ge 4$.

The Einstein-Podolsky-Rosen paradox is a stronger form of entanglement in the sense that it cannot be demonstrated by all mixed entangled states ~\cite{Wiseman2007a}. For our system, we previously showed~\cite{Ferris2008b} that violating
\begin{equation}
  \mathrm{Var}\bigl[\h[\phi_1]{X}_1 - \h[\phi_2]{X}_2\bigr] + \mathrm{Var}\bigl[\h[\phi_1]{Y}_1 + \h[\phi_2]{Y}_2\bigr]
  \ge 2\left| 1 - \frac{\langle\hd{a}_j \h{a}_j\rangle}{\langle\hd{b}_j\h{b}_j\rangle} \right|, \label{EPR1}
\end{equation}
for $j = 1$ or $2$, demonstrates the EPR paradox. This is violated for at least one value of $j$ if $S < 2$.

A similar, yet stronger EPR criterion is
\begin{gather}
  \Delta^2_\mathrm{inf}\bigl[\h[\phi_2]{X}_2\bigr] = \mathrm{Var}\bigl[ \h[\phi_2]{X}_2 \bigr] -
  \frac{\mathrm{Var}\bigl[\h[\phi_1]{X}_1,\h[\phi_2]{X}_2\bigr]^2}{\mathrm{Var}\bigl[\h[\phi_1]{X}_1\bigr]},\nonumber\\
  \Delta^2_\mathrm{inf}\bigl[\h[\phi_2]{Y}_2\bigr] = \mathrm{Var}\bigl[ \h[\phi_2]{Y}_2 \bigr] -
  \frac{\mathrm{Var}\bigl[\h[\phi_1]{Y}_1,\h[\phi_2]{Y}_2\bigr]^2}{\mathrm{Var}\bigl[\h[\phi_2]{Y}_1\bigr]},\label{EPR2}\\
  \mathcal{E} = \frac{\Delta_{\mathrm{inf}}^2\bigl[\h[\phi_2]{X}_2\bigr] \,
  \Delta_{\mathrm{inf}}^2\bigl[\h[\phi_2]{Y}_2\bigr]}{\left( 1 - \langle \hd{a}_j \h{a}_j \rangle/\langle\hd{b}_j\h{b}_j\rangle \right)^2}\ge 1, \nonumber
\end{gather}
where in the last line we defined the EPR parameter $\mathcal{E}$. Minimizing $\mathcal{E}$ with respect to $\phi_1$ and $\phi_2$ is less straightforward than for $\mathcal{S}$, but the two will usually be minimized for similar phase angles\footnote{Generally $\mathcal{E}(\phi_1,\phi_2)$ is minimal at the extreme points of $\mathcal{S}(\phi_1,\phi_2)$, where the (anti-)correlation is strongest.}.

%In the next section we implement a model of our system and demonstrate that each of these entanglement criteria can be violated.

\section{Simulations}

\label{sec_results}

\subsection{Outline of experiment}

We begin with a pure condensate of $N_0 = 10^5$ $^{87}$Rb atoms in a quasi-1D harmonic trap with trapping frequencies $(\omega_z, \omega_{\perp}) = 2\pi\times(1, 44)$ Hz. Due to the high trapping aspect ratio, we assume we can ignore the dynamics in the tightly trapped direction, and take a variational Gaussian ansatz for the transverse wave function. We have chosen system parameters that closely match the previous work of Hilligs{\o}e and M{\o}lmer~\cite{Hilligsoe2005a} to allow for a clear comparison of our results with their calculations. Such a quasi-1D condensate is described by the Lieb-Liniger Hamiltonian with an additional external potential.
\begin{equation}
\h{H} = \int dz\, \hd{\psi}(z)\left( \frac{-\hbar^2}{2m}\frac{\partial^2}{\partial z^2} + V(z,t)  + \frac{g}{2} \hd{\psi}(z)\h{\psi}(z)\right)\h{\psi}(z),
 \label{lieb_liniger}
\end{equation}
where $m$ is the mass of an atom, $g = 4\pi\hbar^2a_s/mA_{\perp}$ is an effective 1D interaction constant, and $V(z,t) = m\omega_{z}^2z^2/2 + V_L(z,t)$, where $V_L$ is the optical lattice potential [Eq.~(\ref{lattice})]. We use a scattering length $a_s = 100 a_0$ and obtain the effective cross-sectional area $A_{\perp} = 42$ $\mu$m$^2$ by minimizing the energy of a Gaussian ansatz, resulting in~\cite{Hilligsoe2005a}
\begin{equation}
  A_{\perp} = \frac{2\pi\hbar}{m\omega_{\perp}} \sqrt{1 + 2a_s \bar{n}_{1\mathrm{D}}},
\end{equation}
where $\bar{n}_{1\mathrm{D}}$ is the average linear density. The condensate ground state wave function is quasi-one-dimensional if $\bar{n}_{1\mathrm{D}} \lesssim a_s^{-1}$. For these parameters the Thomas-Fermi length of the condensate is approximately 295 $\mu$m.

\begin{figure}[t]
\begin{centering}
\includegraphics[width=6.2cm]{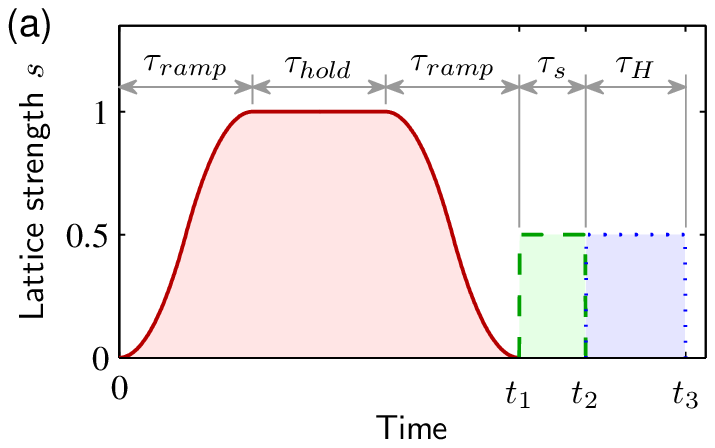}

\vspace{0.1cm}

\includegraphics[width=5.4cm]{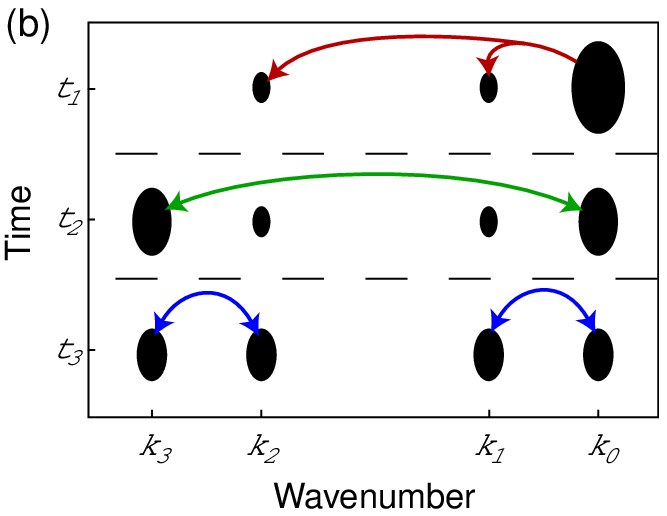}
\caption{(Color online) (a) The intensity of the various optical lattices as a function of time. The solid (red) line indicates the intensity of the optical lattice generating the entanglement, which is ramped on and off with a continuous first derivative using parabolic curves. This process adiabatically transfers momentum to quasi-momentum states and vice-versa. The dashed (green) line indicates the Bragg pulse to split the phase reference in two. The dotted (blue) line indicates the simultaneous Bragg pulses used to beam-split each of modes 1 and 2 with local oscillators in modes 0 and 3. (b) A momentum space schematic of each stage of the process.}
\label{fig_pulses}
\end{centering}
\end{figure}

The experimental sequence to generate and detect entanglement between two matter-wave pulses is illustrated in Fig.~\ref{fig_pulses}
To initiate the degenerate four-wave mixing, an optical lattice of wavelength 790 nm and peak strength $s=1$ is adiabatically ramped on and off over a time $t_1$ as depicted in Fig.~\ref{fig_pulses}(a).  In the simulations we ramp the lattice on and off using piecewise parabolic curves with continuous first derivatives over a time period $\tau_{\mathrm{ramp}} = 2$ ms. Gross-Pitaevskii simulations were performed to ensure that the shape and duration of the ramping curve efficiently transfers momentum states to quasi-momentum states and vice-versa. The optical lattice is detuned by
\begin{equation}
  \delta_0 = -\frac{2 \hbar k_L \Delta\! k}{m},
\end{equation}
where $\Delta\! k = (43/64) k_L$. In the frame of the lattice, the condensate begins with mean momentum $k_0 = \Delta\! k$, as depicted in Fig.~\ref{fig_bands}, and degenerate four-wave-mixing generates new wave packets with mean momenta $k_1$ and $k_2$.

At this point in the simulation collisions between atoms in the system become a hinderance. We therefore assume that the harmonic trap is removed after the optical lattice is switched off. During the subsequent expansion of the atom cloud the collision rate and hence the effective nonlinearity will decrease. Collisions have a negative impact on the measurement scheme and reduce the strength of correlation between the quadratures. If the radial trapping is very strong, releasing the trap suddenly could potentially result in complex three-dimensional collision dynamics that could disturb the fragile entangled state. To avoid most of these collisions, the trapping potential may be ramped down adiabatically over a timescale short compared to the rest of the experiment (but fast compared to $\omega_r$), removing most of the interaction energy from the system while preserving the quasi-one-dimensional state. This leaves the atoms to slowly expand radially in a Gaussian wavepacket. To model this procedure, in the simulations the nonlinear constant $g$ is set to zero at time $t_1$.

The measurement scheme is implemented by applying a series of Bragg pulses. The first pulse is also of wavelength 790 nm, but is tuned to transfer half of the initial condensate at rest (in the lab frame) to mode 3 with momentum $-2k_L$ and occurs between $t_1$ and $t_2$.  Immediately following this between $t_2$ and $t_3$ are two simultaneous Bragg pulses of wavenumber $(k_0 - k_1)/2 = (k_2 - k_3)/2$, each detuned to beam split modes 1 with 0, and modes 2 with 3 respectively, as illustrated in Fig.~\ref{fig_pulses}~(b).

\subsection{Simulation method}

%Solutions to the Lieb-Liniger model without a potential are integrable and can be found using the Bethe ansatz~\cite{Lieb1963,Bloch2008}, however exact solutions to general potentials are difficult to obtain.

We implement the truncated Wigner approximation~\cite{Steel1998,QuantumNoise,Ferris2008} to model the quantum dynamics of the proposed experiment. Despite being approximate, the method is numerically stable and has been shown to accurately treat quantum field dynamics on short to medium timescales. In particular, the validity condition of simulating a large number of particles compared to the number of modes is well satisfied in our numerical calculations~\cite{Blakie2008}.

The truncated Wigner method is implemented by stochastically sampling the initial Wigner functional $W(\psi(z))$ and then evoling this according to the Gross-Pitaevskii equation (GPE)
\begin{equation}
  i\hbar\frac{\partial\psi(z)}{\partial t} = \left(\frac{-\hbar^2}{2m}\frac{\partial^2}{\partial z^2} + V(z,t) + g|\psi(z)|^2\right)\psi(z). \label{gpe}
\end{equation}
The initial state we use is a coherent state condensate found by solving the GPE in imaginary time with the addition of vacuum noise in the remaining empty modes~\cite{Blakie2008}. The ensemble of trajectories then represents the evolution of the Wigner functional. Expectation values of symmetrically ordered quantities are obtained by sampling moments of the field. For instance,
\begin{equation}
  \overline{\psi^{\ast}(z)\psi(z)} = \langle \hd{\psi}(z)\h{\psi}(z) + \h{\psi}(z)\hd{\psi}(z) \rangle / 2.
\end{equation}
We discretize the atomic field into 4096 points with a range of 404~$\mu$m, and the evolution is computed using a split-operator adaptive 9th order Runge-Kutta algorithm generated by the open-source software \texttt{XMDS}~\cite{xmds}.  We typically run $100$--$1000$ trajectories for each set of parameters.

\subsection{Results}

\subsubsection{Maximal entanglement}

We begin with an analysis of simulations of the degenerate four-wave mixing process only [up to time $t_1$ in Fig.~\ref{fig_pulses}(a)] in order to identify the maximum violations of the entanglement inequalities without the complication of the measurement scheme. We employ both the trapped model described above and a simplified, periodic model, that does not include the axial harmonic trap. This allows us to illustrate important differences between the situations with well-defined quasi-momenta, and the more realistic trapped condensate. For the periodic model, we begin with $10^5$ atoms spread over 512 periods of the optical lattice, or 202 $\mu$m, which roughly corresponds to the high density region of the trapped condensate.

\begin{figure}[t]
\begin{centering}
\includegraphics[width=\columnwidth%
]{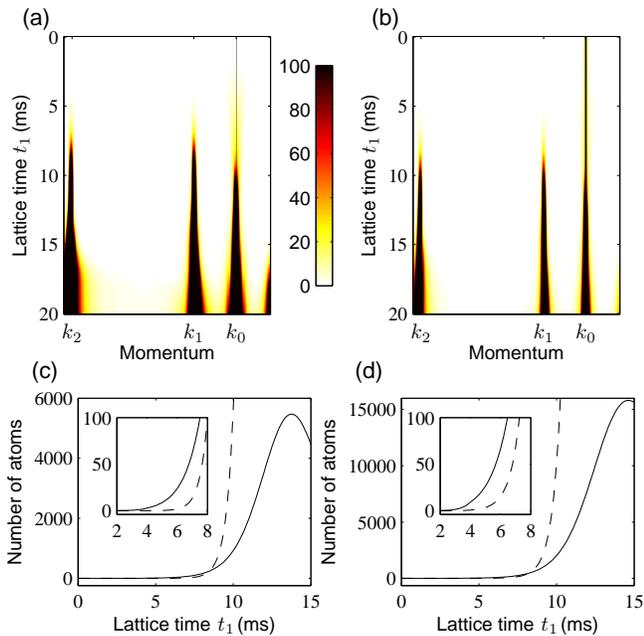}
\caption{(Color online) Momentum density and relative number fluctuations versus the time held in the optical lattice, $t_1$. (a) Population of each discrete momentum mode for the periodic model and (b) the trapped model. The initial condensate is barely visible in the periodic model as occupies just a single momentum mode. We see strong population growth around momenta $k_1$ and $k_2$, as well as low energy scattering about $k_0$. In (c) and (d), the solid line is the number in the entangled modes, $\langle \h{N}_1 + \h{N}_2\rangle$, and the dashed line is the variance of the number difference, $\mathrm{Var}\bigl[\h{N}_1 - \h{N}_2\bigr]$. Relative number squeezing is observed when this variance is less than the total number, and a strong correlation is exhibited at early times in (c), the data from the homogenous model, and (d), the data from the trapped model. At later times the non-classical correlation is lost.}
\label{fig_numbers}
\end{centering}
\end{figure}

In Fig~\ref{fig_numbers}(a,b) we plot the population of atoms in each momentum mode as a function of time held in the lattice, $t_1$. This allows us to determine which momentum modes the degenerate four-wave-mixing process populates most strongly. The results show that the population grows fastest in the modes with momentum $k_0 \pm (105/256)k_L$ modulo $2k_L$ for the periodic model, and $k_0 \pm (104/256)k_L$ modulo $2k_L$ for the trapped model. These values are close to the predictions of the simple band-structure model, depicted in Fig.~\ref{fig_bands}, and differ because of the slight average density differences between the two models. For the remainder of this paper we use values of $(k_0, k_1, k_2, k_3) = (0.672, 0.262, -0.918, -1.381)k_L$ for the periodic model, and $(k_0, k_1, k_2, k_3) = (0.672, 0.266, -0.922, -1.381)k_L$ for the trapped model.

The modes surrounding these peak growth modes also undergo significant growth. The time-energy uncertainty relation allows momenta close to the resonance condition to experience population growth for a timescale inversely related to their detuning. Below, we see that these additional populated modes do not prevent the detection of entanglement. After a longer period of time, secondary collisions populate a range of modes and the condensate is significantly depleted. Previous experimental and theoretical work has shown this behavior is a result of the well-known dynamical instability present in moving optical lattices~\cite{WuNuBattle,Konotop2002,Ferris2008}. Instabilities in optical lattices have traditionally been studied due to their thermalizing effects; however the degenerate four-wave-mixing process we employ to create entanglement \emph{is} a dynamical instability in mean-field terminology.

In Ref.~\cite{Hilligsoe2005a}, Hilligs{\o}e and M{\o}lmer~ implement a mean-field model and predict up to 95\% transfer of the population to modes 1 and 2 (when an initial population seed is placed in mode 1). Our simulations account for spontaneous collisions into many modes and show that the transfer efficiency is actually limited to much less than this value without seeding. In this situation for the trapped case we reach a maximum of 16\% population transferred to the windows about $k_1$ and $k_2$, and with a 10\% seed we get a total of 45\% conversion, showing that spontaneous scattering into multiple modes limits the transfer efficeincy.

We have calculated the relative number squeezing between the two signal modes centered about momentum $k_1$ and $k_2$. In Fig.~\ref{fig_numbers}~(c,d) we see the population in each mode grows approximately exponentially as one would expect for this system~\cite{Olsen2006a}.
It is important to note that a trapped condensate has a nonzero momentum spread. Detecting only a narrow momentum range of the trapped condensate would result in a poor overlap with the true spatial mode, degrading both the number correlation and the performance of the entanglement criteria. One must make measurements over a range of momenta of width $2\delta k$. The number measured in signal mode $j$ is then
\begin{equation}
 \h{N}_{j} = \int_{- \delta k}^{+\delta k} \hd{\psi}(k_j + k^{\prime})\h{\psi}(k_j + k^{\prime}) \, dk^{\prime}.
\end{equation}
In Appendix~\ref{app_multimode} it is shown that the entanglement criteria described in Eqs.~(\ref{sep_param}-\ref{EPR2}) hold true provided $\delta k$ is small enough, replacing $\langle \hd{a}_j \h{a}_j \rangle$ with $\langle\h{N}_j\rangle$.

\begin{figure}[t]
\begin{centering}
\includegraphics[width=8cm%\columnwidth
]{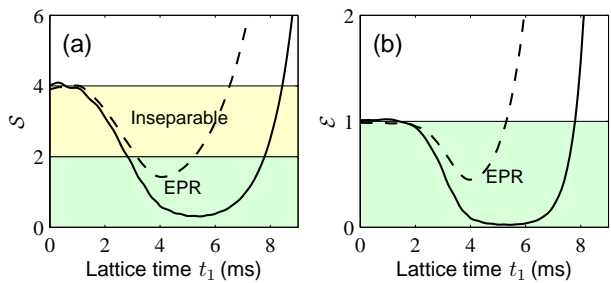}
\caption{(Color online) Entanglement demonstration as a function of the hold time in the optical lattice, $t_1$. The solid line represents the result for the homogenous calculation, while the dashed line is the result for the trapped calculation. (a) The separability criterion [Eq.~(\ref{Duan2})] is maximally violated when the entangled populations $N_1 + N_2 \approx 9$. The presence of the trap reduces the level of violation, but the system still demonstrates the EPR paradox as given by Eq.~(\ref{EPR1}). (b) The two systems display similar behavior to (a) with respect the EPR criterion, Eq.~(\ref{EPR2}). Ensembles of 1000 stochastic trajectories were used to generate these results.}
\label{fig_ideal}
\end{centering}
\end{figure}

In Fig.~\ref{fig_ideal}~(c,d) we sum over five modes in the computational basis to determine the number of atoms in each pulse. We also see the number difference variance between the two signal modes is significantly below the shot noise level $\langle \h{N}_1 + \h{N}_2\rangle$ for smaller values of $t_1$. The number squeezing degrades at later times due to secondary collisions transferring atoms out of the two signal pulses (in agreement with~\cite{Ogren2008}).

We now analyze the entanglement of the modes 1 and 2 by implementing an idealized measurement scheme. In Fig.~\ref{fig_ideal} we plot the results of direct measurement of the quadrature operators described by Eqs.~(\ref{quad1},\ref{quad2}). The separability parameter $\mathcal{S}$ and EPR parameter $\mathcal{E}$ have been minimized with respect to the quadrature angles $\phi_1$ and $\phi_2$. To do this we directly extract from the simulations the expectation values in the expression Eq.~(\ref{sep_angle}), and from these determine the phase-angle $\phi_1+\phi_2$ which minimizes $\mathcal{S}$. We find that the optimal phase-angle varies significantly with time.  For simplicity, we assume this phase-angle also minimizes $\mathcal{E}$, which we find to be true in practice. Entanglement and the EPR paradox are indeed demonstrated for some time in both models. After a few milliseconds the quantum state of the field becomes complicated, and the entanglement detectable by these separability and EPR criteria disappears in agreement with previous work~\cite{Olsen2006a,Ferris2008b}.

%We note that the entanglement criteria in Fig.~\ref{fig_ideal} are slightly underestimated due to the estimator bias. To obtain these results we have minimized with respect to the quadrature angles, naively ignoring the fact that we use a finite number of samples. The estimator bias of optimization problems can be significant as it may scale poorly with respect to sample size.

\begin{table}
\caption{The maximal amounts of separability and EPR violation predicted for periodic and trapped systems beginning with different initial seed populations.\label{tab_rates}}
\begin{center}
\begin{ruledtabular}
\begin{tabular}{ccccrr}
  \begin{minipage}[b]{0.8cm} Trap \end{minipage} &
  \begin{minipage}[b]{0.9cm} Seed size \end{minipage} &
  \begin{minipage}[b]{1.6cm} Number in entangled modes \end{minipage} &
  \begin{minipage}[b]{1.4cm} Number difference variance \end{minipage} &
  \begin{minipage}[b]{1.6cm} Separability violation \end{minipage} &
  \begin{minipage}[b]{1.5cm} EPR violation \end{minipage} \\
  \hline
  No & $0$ & $13$ & $0.5$ & $11.1$ dB & $16.2$ dB\\
  No & $10$ & $149$ & $16$ & $10.9$ dB & $15.8$ dB\\
  No & $10^2$ & $936$ & $124$ & $9.6$ dB & $13.4$ dB\\
  No & $10^3$ & $3920$ & $1053$ & $5.6$ dB & $5.9$ dB\\
  No & $10^4$ & $10465$ & $10243$ & $-0.1$ dB & $-2.2$ dB\\
  Yes & $0$ & $9.3$ & $0.7$ & $4.5$ dB & $3.5$ dB\\
  Yes & $10$ & $38$ & $11$ & $4.5$ dB & $3.5$ dB\\
  Yes & $10^2$ & $296$ & $103$ & $4.4$ dB & $3.3$ dB\\
  Yes & $10^3$ & $2270$ & $986$ & $3.7$ dB & $2.1$ dB \\
  Yes & $10^4$ & $9963$ & $9836$ & $-0.1$ dB & $-2.1$ dB
\end{tabular}
\end{ruledtabular}
\end{center}
\end{table}

We include the magnitude of the maximal entanglement criteria violation (with respect to $t_1$) in Table~\ref{tab_rates} along with the number of entangled atoms. In both models the peak violation occurs when the total population in modes 1 and 2, $N_1 + N_2$, is approximately 10. The turning points of $\mathcal{S}$ and $\mathcal{E}$ correspond to when the number difference variance begins to grow. We note that the simplified, periodic system produces as much as 10 dB stronger inequality violations compared to the trapped case.

\begin{figure}[t]
\begin{centering}
\includegraphics[width=8cm%\columnwidth
]{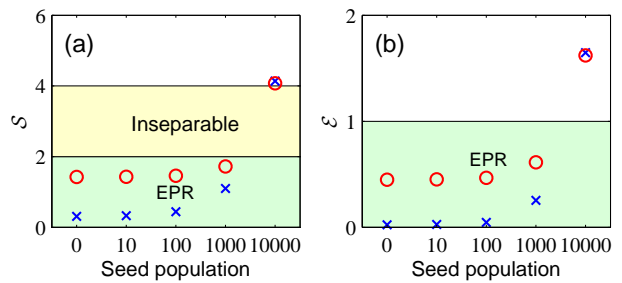}
\caption{(Color online) Minimal values of (a) the separability parameter $\mathcal{S}$ and (b) the EPR parameter $\mathcal{E}$ with different initial seed populations in mode 1. Crosses (blue) represent results for periodic systems and circles (red) for trapped systems. Larger seeds appear to degrade the amount of entanglement violation. Each point corresponds to a different time $t_1$ that minimises the separability parameter for each system and seed size.}
\label{fig_seed}
\end{centering}
\end{figure}

\subsubsection{Seeding signal modes}

In optical experiments bright entangled sources can be created by `seeding' the four-wave-mixing process with a coherent input into one of the signal modes~\cite{Zhang2000}. We have investigated populating mode 1 with a seed in order to boost the number of atoms in the entangled modes. This could be achieved experimentally with a short Bragg pulse before the optical lattice is applied.

Figure \ref{fig_seed} and Table~\ref{tab_rates} indicate the results of our analysis. In common with the optical case, we find that the entanglement violation is decreased for larger seeds. A seed of even 1\% of the initial population, or 1000 atoms, results in noticeable degradation, and a 10\% seed leads to no entanglement criteria violation at any time. We believe increased secondary scattering is the cause of this degradation. Despite the fact that larger signals can be obtained by using a seed, the useable entanglement, such as measured by the entropy of entanglement~\cite{Braunstein2005}, does not increase with a larger seed\footnote{This can be seen from Eq.~(\ref{upa_solution}); if one adds a coherent seed in mode 1 by a displacement operation, $\h{a}_1 \rightarrow \h{a}_1 + \alpha$ in the Heisenberg picture, then the resulting solutions are also simply displaced, and the entropy of entanglement is unchanged.}.

\subsubsection{Implementation of measurement scheme}

\begin{figure}[t]
\begin{centering}
\includegraphics[width=8cm%\columnwidth
]{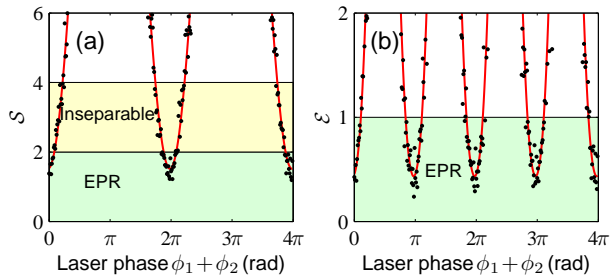}
\caption{(Color online) The measured dependence of the (a) separability and (b) EPR parameters on the phase of the Bragg lasers $\theta_1 - \theta_2 = \phi_1 + \phi_2$. Dots (black) indicate data from stochastic simulations for each phase angle. Solid (red) lines represent fits to the data of (a) sinusoidal form with minimum value $\mathcal{S} = 1.52$ and (b) two-frequency sinusoidal form with minima of $\mathcal{E} = 0.43$. These results are directly calculated from number correlations at time $t_3$.}
\label{fig_measure}
\end{centering}
\end{figure}

Finally, we have performed simulations implementing the sequence of Bragg pulses that are required to demonstrate entanglement in our scheme. To reiterate, first the optical lattice is applied to generate the entanglement until time $t_1$. Immeditately following this we set $g$ and $\omega_z$ to zero (representing rapidly removing the trap on a timescale quick compared to $1/\omega_z$ but short compared to $1/\omega_r$) and begin the Bragg pulse splitting the condensate into modes $0$ and $3$. This pulse ends at time $t_2$, when we apply two superimposed $\pi/2$ Bragg pulses which mix modes 0 and 1, and modes 2 and 3. At time $t_3$ the quadratures are measured by the appropriate number differences between the modes in $k$-space. During the time the pulse sequence is applied the four separate condensates are overlapping in real space.  To perform the homodyne measurements it would be necessary to allow these to expand and separate in real space, and then then accurately count the number of atoms in each separate cloud. Note that a heterodyne scheme may be possible by investigating the real-space spectral components of the density at time $t_3$, but may be difficult with limited detector resolution.

We have carried out independent simulations of 360 different Bragg pulse phase settings with $\theta_1 = -\theta_2$. We find the optimal angle for violating the entanglement criteria in the same manner as would be necessary in an experiment. The results are plotted in Fig.~\ref{fig_measure}, where each data point is the result of an ensemble of 100 stochastic trajectories. The statistical variations of these points give some indication of what may be expected from a similar number of experimental runs\footnote{Although single and ensembles of truncated-Wigner trajectories are not formally equivalent to the expected results of experimental realizations, there is some degree of correspondence \cite{Blakie2008}.}. Previous statistical experiments of BECs have involved tens~\cite{Folling2005a} to thousands~\cite{Schellekens2005a,Vassen2007a} of shots.

We have performed a sinusoidal fit to the data to determine the maximal separability and EPR violation values. The separability criteria follows a simple sinusoidal form [Eq.~(\ref{sep_angle})] with a minima of 1.52, or a 4.2 dB violation. The analytic form for the EPR parameter as a function of quadrature angle is more complicated; we fit the sum of two sinusoids (with periods $\pi$ and $\pi/2$) resulting in the fitted minimum value of 0.43, or a violation of 3.8 dB. We see that these values fit closely to those of the idealized measurement scheme (c.f. Table~\ref{tab_rates}), and that the measurement scheme as simulated is highly efficient.  Of course technical issues such as detector efficiencies in a real experiment will potentially have a significant impact on the results.

\section{Conclusions}

\label{sec_conclusion}

We have combined the entanglement generating method of degenerate four-wave mixing for a trapped Bose-Einstein condensate loaded into a moving optical lattice proposed by Hilligs{\o}e and M{\o}lmer~\cite{Hilligsoe2005a} with a quadrature measurement scheme and entanglement criteria previously derived in Ref.~\cite{Ferris2008b}.  Once the entanglement is generated, a series of Bragg pulses are applied in analogy to beam splitting used by homodyne schemes in quantum optics. Finally, the number of atoms in each mode is measured via time-of-flight expansion, and the quadrature values are determined from these.

We have simulated a numerical model of a realistic quasi-one dimensional condensate that incorporates the effects of imperfect beam splitting by Bragg pulses, non-adiabaticity of the optical lattice, and the effects of the trapping potential. Our results indicate that neither multimode effects or the presence of a trap will necessarily prevent the demonstration of entanglement with ultra-cold atoms. A theoretical analysis of different atom types, geometry and other experimental parameters may lead to significant optimizations and improvements to the predicted entanglement criteria violations.

These results are promising, but it is important to note potential experimental difficulties that we have neglected in our model. Most significantly, detector efficiency plays a crucial role in entanglement demonstration. Atom detection techniques are currently undergoing rapid development and we expect that modern detectors with efficiencies of 30--50\% should be sufficient to demonstrate inseparability with this scheme~\cite{Bowen2004}. It is critical that the phase of the Bragg pulse and optical lattice lasers be carefully controlled, in order to measure the quadrature statistics of a specific quadrature angle. The effects of three-body loss and finite-temperature effects should also be taken into consideration. We have assumed the problem is purely one-dimensional; one may expect collisions into higher energy radial spatial modes during the four-wave mixing and the expansion process. These collisions can be reduced with tighter radial trapping frequencies, but truly 1D systems suffer from phase fragmentation which may have other detrimental effects.

Despite these potential complications we believe that we have outlined a feasible scheme for demonstrating entanglement between modes of ultra-cold bosonic atoms. We would like to point out that a similar detection scheme could be applied to entanglement generated in a different manner, such as by colliding two BECs in free space~\cite{Deng1999,Vogels2002,Perrin2007,Perrin2008,Ogren2008}. However, we remind the reader that the advantage of the quasi-1D geometry of the scheme we describe is it that limits the number of resonant modes and results in only two entangled matter-wave packets. It is possible to imagine other schemes that similarly populate few entangled modes; for instance a recent experiment observed  well-separated and potentially entangled modes generated by four-wave mixing between different hyperfine states of metastable helium~\cite{Dall2009}. Not only is entanglement between massive particles interesting in its own right, but producing entangled atomic sources may lead to further novel experiments and quantum information applications in the future.

\acknowledgments

The authors acknowledge financial support from the Australian Research Council Centre of Excellence program.

\appendix
\section{Multiple modes}

\label{app_multimode}
It is straightforward to extend the entanglement criteria of Ref.~\cite{Ferris2008b} to the situation where the signal modes have a finite momentum width. Realistic experiments will involve condensates with uncertain momentum, undergoing finite time-of-flight before measurement on detectors with non-zero resolution.

We consider a range of momentum values small enough that the Bragg pulses act similarly for each momenta. We write the number measured about each relevant momenta $k_j$ as
\begin{equation}
 \h{N}_{j}(t) = \int_{- \delta k}^{+\delta k} \hd{\psi}\h{\psi}(k_j + k^{\prime}, t) \, dk^{\prime},
\end{equation}
and then measure each quadrature similarly to the single mode case. The numbers after the final Bragg pulse relate to the quadrature before; for example,
\begin{eqnarray}
  \h{N}_1(t_3) - \h{N}_0(t_3)\! & = \!& \!\int_{- \delta k}^{+\delta k} \!\!\!\!\!\!\! \hd{\psi}(k_0 + k^{\prime}, t_2)\hd{\psi}(k_1 + k^{\prime}, t_2)e^{-i\phi_1}  \nonumber \\
 & & \!+ \hd{\psi}(k_1 + k^{\prime}, t_2)\h{\psi}(k_0 + k^{\prime}, t_2)e^{i\phi_1} \, dk^{\prime} \nonumber \\
 & \equiv & \langle \h{N}_0(t_2) \rangle^{1/2} \h[\phi_1]{X}_1(t_2).
\end{eqnarray}
The above assumes $\delta k$ is within the range of values that the Bragg pulse affects, given by the time-energy uncertainty principle, $\hbar \, \delta k^2 / 2m \lesssim 1 / \tau_H$.

We can see that commutation relations are very similar to those in Eq.~(\ref{commutator}),
\begin{equation}
   \left\langle \left[ \h[\phi]{X}_1, \h[\phi]{Y}_1 \right]\right\rangle = 2i \left(1 - \frac{\langle \h{N}_1(t_2) \rangle }{ \langle \h{N}_0(t_2) \rangle } \right).
\end{equation}
It follows that the entanglement criteria in Eqs.~(\ref{Duan2},\ref{EPR1},\ref{EPR2}) hold true, replacing $\langle \hd{a}_j \h{a}_j \rangle$ with the total number in the momentum range, $\langle \h{N}_j \rangle$.

%\bibliography{andy}
%\bibliography{../../bib/andy}
%\bibliographystyle{apsrev}

\end{document}